\documentstyle[psfig,aps,prl,multicol]{revtex}

\renewcommand{\narrowtext}{\begin{multicols}{2} \global\columnwidth20.5pc}
\renewcommand{\widetext}{\end{multicols} \global\columnwidth42.5pc}

\begin{document}
\title{Reaching the continuum limit in lattice gauge theory - without a computer.}
\author{John A.L. McIntosh and Lloyd C.L. Hollenberg}
\address{Research Centre for High Energy Physics\\
School of Physics\\
The University of Melbourne\\
Victoria 3010, Australia\\
(j.mcintosh@physics.unimelb.edu.au, l.hollenberg@physics.unimelb.edu.au)}
\maketitle
\begin{abstract}
The scaling slope of the anti-symmetric mass gap $M$ of compact U(1)$_{2+1}$ lattice
gauge theory is obtained analytically in the Hamiltonian formalism
using the plaquette expansion. Based on the first four
moments of the Hamiltonian with respect to a one-plaquette mean
field state the results demonstrate clear scaling of $M$ at and
beyond the transition from strong to weak coupling. The scaling
parameters determined agree well with the range of numerical
determinations available. 

\vspace*{0.2cm}

\end{abstract}

\narrowtext

With the vast majority of calculations carried out in lattice
gauge theory being of an intensely numerical nature,
non-perturbative analytic results are of considerable interest. If
we look to complex theories such as QCD, the most basic
quantities one wishes to calculate are the states of the mass
spectrum of the theory. In lattice gauge theory, such
non-perturbative calculations are carried out numerically in the
path integral representation to some level of approximation by
evaluating the correlation functions by Monte-Carlo simulation.
Typically, one must work with large lattices and spacings which
are fine enough to detect the continuum behaviour of the theory in
question. The computational resources employed in this approach
are enormous, essentially reflecting the seemingly intractable
many-body problem at the core of non-perturbative quantum field
theory. We report here on a novel approach to studying lattice
gauge theories, and as an indication of its potential apply the
method to compact U(1) lattice gauge theory in 2+1 dimensions.
This theory, although obviously simpler than QCD itself,
nevertheless shares many of the features of the QCD many-body
problem on the lattice, namely similar numerical difficulties in
calculating masses in the continuum limit. 
The scaling behaviour
for the mass gap, $M$, is expected~\cite{gopmack82} to be:
\begin{equation}
M^{2} a^{2}\equiv\beta e^{-k_{0}\beta\ +\ k_{1}}\ ,
\label{scale}
\end{equation}
where $k_{0}$ and $k_{1}$ are constants, $a$ is the lattice spacing and
$\beta =1/g^{2}$ where $g$ is the dimensionless coupling constant.
As with most other
interesting lattice gauge theories, it has long been a maxim that
even U(1)$_{2+1}$ lattice gauge theory must be studied by large
scale numerical computation in order to determine the lowest mass
state of the theory. 
Indeed, although studied by a range of
different methods, the scaling properties of this theory are still
not known precisely: 
$k_{0}\sim 4.1 - 6.345$ and $k_{1}\sim 4.369 - 6.27$~(see 
\cite{inprogress} for summary).
In this sense the
results we present are quite spectacular -- an analytic
determination of the mass-gap, which clearly demonstrates the
scaling behaviour expected in the continuum limit with $k_{0}=4.42$
and $k_{1}=5.41$.

Throughout this paper we will work in the Hamiltonian formalism.
When one speaks of ``lattice gauge theory'' it is usually meant
the Lagrangian formalism where one essentially converts the
theory to a statistical model in Euclidean space into order to
facilitate the use of powerful Monte-Carlo methods. In contrast,
there is a dearth of equivalently efficient methodology in the Hamiltonian formalism of
lattice gauge theory. This fact has been responsible for the
domination of action-based lattice gauge theory over the
Hamiltonian approach. While there exist numerical and Monte-Carlo
methods applicable to the Hamiltonian lattice formalism, these
have been of limited use, and certainly the results are not of the
same quality as in the action-based approach. However, there is a
certain appeal of the Hamiltonian formalism which tugs at
physicists' intuition. There are also technical reasons
for considering the Hamiltonian approach - one often works
directly on infinite lattices, fermions are simpler to incorporate
(no fermion determinant), the lattices are 3D as opposed to 4D (time is
continuous), and finite temperature physics may actually be more
natural in this formalism\cite{kroeger}. The lack of powerful and systematic
methodology, such as is available in the
Lagrangian formalism, has prevented these features from being
exploited or investigated - hence, the importance of new
methodologies. Here we use a novel 
approach -- the plaquette expansion -- based on a large volume
expansion of Lanczos tridiagonalization.

The Kogut-Susskind Hamiltonian for U$(1)_{2+1}$ on a lattice
of $N$ plaquettes is given by~\cite{kogsuss75}:
\begin{equation}
H=\frac{g^2}2\sum_{l}{\hat E}^2_{l}
 + \frac1{2g^2}\sum_{p}\left[2-({\hat U}_{p} + {\hat U}_{p}^{\dag})\right]\ ,
\label{eq:hamu1}
\end{equation}
The strong-coupling limit is defined by~$g\rightarrow\infty$ and
the weak-coupling limit by $g\rightarrow 0$. For the U(1) model,
${\hat E}_{l}$ is the electric field operator associated with
lattice link $l$. In the representation in which ${\hat E}_{l}$ is
diagonal, the eigenvalue $E_{l}$ takes on integer values $0,\pm\
1,\pm\ 2,\ldots$ on each link and the link operators ${\hat
U}_{l}$ and ${\hat U}_{l}^{\dagger}$ raise or lower the value of
$E_{l}$ on link $l$ by unity respectively due to the 
commutation relations:
\begin{equation}
[{\hat E}_{l},{\hat U}_{l}]={\hat U}_{l}\ \ \ \ \ ,\ \ \ \ \
[{\hat E}_{l},{\hat U}_{l}^{\dag}]=-{\hat U}_{l}^{\dag}\ .
\end{equation}
\noindent The operator ${\hat U}_{p}$ acts on the links around the
smallest closed (Wilson) loop or square on the lattice known
commonly as a plaquette $p$:
\begin{equation}
{\hat U}_{p}={\hat U}_{1}{\hat U}_{2}{\hat U}_{3}^{\dag}{\hat
U}_{4}^{\dag}\ .
\end{equation}

The extensive nature of the Hamiltonian Equation~(\ref{eq:hamu1})
lends itself to the application of the plaquette expansion. Indeed
this method was originally devised to study such local lattice
models\cite{holl93}. Beginning with the well known Lanczos
recurrence for the iterative transformation of the Hamiltonian
into tri-diagonal form, a trial state $|\psi_{1}\rangle$ with the
desired symmetries of the state of interest $|\Psi_{0}\rangle$
(the ground state or vacuum in this case) and non-zero overlap
({\it i.e.} $\langle\psi_{1}|\Psi_{0}\rangle\neq 0$) is chosen and
a basis $\{|\psi_{n}\rangle\}$ is constructed according to
\begin{equation}
|\psi_{n}\rangle=\frac{1}{\beta_{n-1}}
\left[\left(H-\alpha_{n-1}\right)|\psi_{n-1}\rangle
-\beta_{n-2}|\psi_{n-2}\rangle\right],
\end{equation}
where $\alpha_{n}=\langle\psi_{n}|H|\psi_{n}\rangle$ and
$\beta_{n}=\langle\psi_{n+1}|H|\psi_{n}\rangle$ are the matrix
elements of the Hamiltonian in tri-diagonal form.

The matrix elements, $\alpha_{n}$ and $\beta_{n}$,
 can be written in terms of Hamiltonian moments $\langle
H^{n}\rangle\equiv\langle\psi_{1}|H^{n}|\psi_{1}\rangle$. As the
connected part of the Hamiltonian moment is proportional to the
volume of the system $\langle H^{n}\rangle_{c}=e_{n}\,N$, one may
utilize the cumulant transformation \cite{hornwein84},
\begin{equation}
\langle H^{n}\rangle_{c}=\langle H^{n}\rangle-\sum^{n-2}_{p=0}
 {n-1\choose p}
\langle H^{p+1}\rangle_{c}\langle H^{n-1-p}\rangle\ ,
\label{eq:hw84}
\end{equation}
and rewrite the $\alpha_{n}$ and $\beta_{n}$ in terms of the
connected vacuum coefficients $e_{n}$ . Extensivity of the problem
leads to the following plaquette expansions in $1/N$\cite{holl93}:
\begin{eqnarray}
{\alpha_n\over N}&=& e_{1} + s\left[\frac{e_{3}}{e_{2}}\right] +
O(s^2)\equiv {\bar \alpha}(s), \label{eq:avac}
\\ {\beta_n^{2}\over N^2} &=&s e_{2} + \frac12
s^{2}\left[\frac{e_{2}e_{4}-e^2_{3}}{e^2_{2}} \right] +
O(s^3)\equiv {\bar \beta^2(s)}, \label{eq:bvac}
\end{eqnarray}
where $s\equiv n/N$. In the bulk limit ($n,N\rightarrow\infty$),
keeping $s$ fixed, it was shown in~\cite{hwaf95} that
one may perform  the exact diagonalization of
the Lanczos tri-diagonal matrix for the vacuum energy
density $\varepsilon_{0}$:
\begin{equation}
\varepsilon_{0}= {\rm inf}\, ({\bar \alpha}(s) - 2{\bar
\beta(s)})\, \label{eq:f0}
\end{equation}
For low order truncations of the series (employing moments up to
$\langle H^{r}\rangle$) one can minimize (\ref{eq:f0}) with respect to $s$
analytically and obtain the closed form expressions for the ground
state energy, $\varepsilon_{0}(r)$:
\begin{eqnarray}
\varepsilon_{0}(1) &=& e_1,\cr
\varepsilon_{0}(3) &=& e_1 - {e_2^2\over e_3},\cr
\varepsilon_{0}(4) &=& e_{1} + \frac{e_{2}^{2}}{e_{2} e_4 -
e_{3}^{2}}\,\left[ \sqrt{3 e_{3}^{2}-2 e_{2} e_{4}}-e_{3}\right].
\label{eq:E0}
\end{eqnarray}
Note the first approximant $\varepsilon_{0}(1)$ is just the
variational result. The expressions for $\varepsilon_{0}(3)$ and
$\varepsilon_{0}(4)$ clearly illustrate the non-perturbative
nature of the plaquette expansion as an effective summation over
clusters of increasing complexity.

To determine expressions for excited states we note that lowest
state of a given sector, {\it S}, in the spectrum will in general
have an energy:
\begin{equation}
E^{(S)}_{0}=\varepsilon_{0}\,N+ M^{(S)}_{0}\ . \label{eq:M0}
\end{equation}
where $M^{(S)}_{0}$ is the mass gap. For excited states the
Hamiltonian moments have the form $\langle
H^{n}\rangle^{(S)}_{c}=e_{n} N + m_{n}^{(S)}$. Thus, we can then
define analogous ``cumulants'' $e^{(S)}_{n}$ for this
sector of Hilbert space as $e^{(S)}_{n}\equiv e_{n}+m^{(S)}_{n}/N$,
where we call $m^{(S)}_{n}$ the mass gap connected moments. One
can derive expressions for approximants, $M^{(S)}_{0}(r)$, to the
mass gap, in a similar fashion to those for the vacuum 
energy density. For example, using up to 4th order moments we
have\cite{hollwilwit95}:
\begin{equation}
M^{(S)}_{0}(4)=\sum^{4}_{n=1}\,m^{(s)}_{n}\,e_2^{n-1}\,F_{n},
\label{eq:mfmg}
\end{equation}
where the vacuum moment functions, $F_{n}$, are given by
{\small
\begin{eqnarray}
F_1 &=& 1,\nonumber\\
F_2 &=& {{\,{\sqrt{\Delta}}\,
       \left( \Omega \right)  +
      e_3\,\left( -\Omega +\,e_2\,{{e_3}^2}\,e_4 -\,{{e_2}^2}\,{{e_4}^2} \right)}\over
    {\left( \Delta \right) \,
      {{\left( {{e_3}^2} - e_2\,e_4 \right) }^2}}},\nonumber\\
&&\nonumber\\
F_3 &=& {
      \,\left(2\,{{e_2}^2}\,{{e_4}^2} -3\,{{e_3}^4}
- e_2\,{{e_3}^2}\,e_4 \right) +{\,e_3\,{\sqrt{\Delta}}\,
       \left( 3\,{{e_3}^2} - e_2\,e_4 \right)}\over
    {\left( \Delta \right) \,
      {{\left( {{e_3}^2} - e_2\,e_4 \right) }^2}}},\nonumber\\
&&\nonumber\\
F_4 &=& {
      \,{\sqrt{\Delta}}\,
       \left(e_2\,e_4 - 2\,{{e_3}^2} \right)
+ {\,e_3\,\left( \Delta \right) }\over
    {\left( \Delta \right) \,
      {{\left( {{e_3}^2} - e_2\,e_4 \right) }^2}}},
\end{eqnarray}
}
\noindent with $\Delta=3\,{{e_3}^2} - 2\,e_2\,e_4$ and
$\Omega=8\,e_2\,{{e_3}^2}\,e_4 - 6\,{{e_3}^4} - 3\,{{e_2}^2}\,{{e_4}^2}$.

It is appropriate to comment on relevant methods, which by and
large employ trial states encompassing as much of the expected
continuum physics as is viable. Often, such calculations have
employed the simplest gauge invariant trial state --- the
strong-coupling vacuum $|0\rangle$ (defined as the state
satisfying ${\hat E}_{l}|0\rangle =0$ for every link). This is the
perturbative starting point for series calculations, which are
then extrapolated to weak coupling. Although the calculation of
moments with respect to this state can be carried out to
relatively high order, typically, one finds that this state is
simply inadequate to explore the non-perturbative weak-coupling
regime of the theory. In the case of the plaquette expansion, a
window of scaling in terms of the expansion order was evident, but
higher order results became problematic~\cite{macholl97}.

More recently, trial states which are more complex and appropriate
to weak-coupling have been investigated by a number of authors
\cite{lang84,heysstump85,bish96,bish97,dunp1,duncan} in varying
contexts. There is a clear tradeoff between physical complexity of
the trial state, and algebraic complexity of the moment
calculation. Given this, we utilized a one plaquette exponential
trial state, that has the correct weak coupling behaviour for the
vacuum energy density -- the so-called {\it mean-field} state --
defined as:
$|\psi_{1}\rangle=\mbox{exp}\{{\hat S}/2\}|0\rangle, \label{eq:mfgs}\;$
where
${\hat S} = \lambda\,\sum_{p}\{{\hat U}_{p}+{\hat U}_{p}^{\dagger}\}\;$
and $\lambda$ will eventually be determined variationally from
$\langle H\rangle$. It is convenient to work in the so-called $U$
representation where the trial state is a function of $U_p$
variables:
\begin{equation}
\psi_1[U_p] = e^{S[U_p]/2} = \exp\left\{\lambda
\sum_{p}\left\{{U}_{p}+{U}_{p}^{\dagger}\right\}\right\}
\end{equation}
As the state is constructed from plaquette variables, $U_{p}$, it
is automatically gauge invariant. One expects the true ground
state of the system to have contributions from all possible Wilson
loops -- an important feature of the plaquette expansion method is
that by its very nature larger sized Wilson loops (larger
clusters) are systematically introduced.

In $D=2+1$ the Hamiltonian moments can be written in terms of
integrals over plaquette variables, for example the vacuum moments
are given schematically by:
\[
\langle H^n\rangle = {\int{\cal D}U_p \,\psi_1[U_p]^\dagger \,H^n
\,\psi_1[U_p]\
                   \over \int{\cal D}U_p \,\psi_1[U_p]^\dagger \,\psi_1[U_p]}
\equiv {\int{\cal D}U_p \,G_n[U_p]\,{\rm e}^{S[U_p]}\over \int{\cal
D}U_p \,{\rm e}^{S[U_p]}}
\]
For the group U(1), the plaquette variables may be rewritten in
terms of plaquette phase angles
$U_{p}\equiv\mbox{exp}(i\,\phi_{p})$. The calculation of
Hamiltonian moments for use in the plaquette expansion utilizing
the one-plaquette state in $2+1$ dimensions comes down to
evaluating integrals of the following form,
\begin{equation}
\langle H^{n}\rangle\equiv \left[{1\over { 2
\pi\,I_{0}(2\lambda)}}\right]^{N}
\int^{\pi}_{-\pi}\,\prod_{p}\,d\phi_{p}\,
\,\,G_n\left[\phi_{p}\right]\,\mbox{e}^{S[\phi_p]},
\end{equation}
where $I_{0}(2\lambda)$ is the zeroth-order modified Bessel
function of the first kind. The integrand functions,
$G_n[\phi_{p}]$, naturally become more complex as one increases
the moment order, however, all integrations can be carried out
analytically. All expressions are analytic in $\lambda$ and
$\beta$, and once $\langle H\rangle$ is minimized we implicity
determine $\lambda(\beta)$. 

It is well known \cite{bish96,bish93} that the variational
calculation $\langle H\rangle$ applying the one-plaquette state in
$D=2+1$ dimensions involves no correlations between plaquettes and
thus one is simply solving a Mathieu-type problem. In this
context, the benefit of choosing the one-plaquette state as the
trial state rather than the strong-coupling vacuum becomes more
apparent on investigation of the strong ($\beta \!\rightarrow\!
0$) and weak-coupling ($\beta\!\rightarrow\!\infty$) limits of
$\langle H\rangle$. For the strong-coupling limit, one finds for
the energy density:
$\varepsilon_{0}(1)=\langle H\rangle/\mbox{N}\sim\beta-\beta^{3}/4
+\mbox{O}\!\left(\beta^{7}\right),
\;$
which agrees with strong-coupling perturbation theory~\cite{irvham84a}. 
Similarly, for the weak-coupling limit, one finds:
$\varepsilon_{0}(1)\sim
A - A^{2}/(8\beta) + \mbox{O}\!\left(1/\beta^{2}\right),
\;$
with $A=1$. This agrees exactly with weak-coupling perturbation theory for
the Mathieu problem to the order shown. Exact $\mbox{U}(1)_{2+1}$ has a
similar weak-coupling expansion \cite{dabring91,hamzheng93} with
$A=0.9581$. With the inclusion of larger sized Wilson loops
obviously the U(1) calculations, as they should, will deviate from
the Mathieu results. As this trial state encompasses
the actual behaviour in both 
limits, one has confidence in using the plaquette expansion to
systematically obtain results throughout the range of $\beta$.

\begin{figure}
\vspace{0mm}
\hbox{\hspace{0mm}\parindent=0mm
        \vtop{%
                \psfig{file=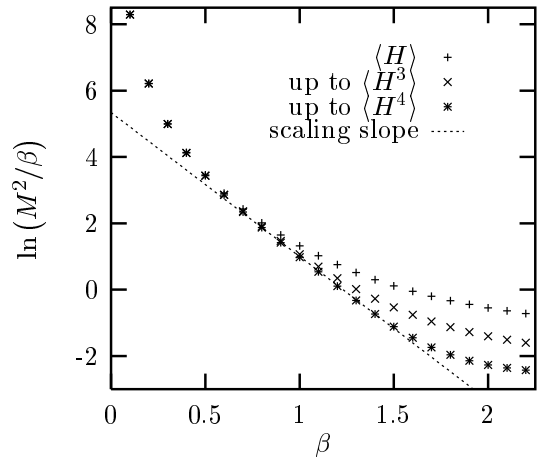,width=70mm}
                \vbox{\hsize=70mm}
        }
}
\vspace{-0.2cm}
\caption{Mass gap results applying one (minimized numerically), three and four one-plaquette
Hamiltonian moments. Scaling is evident for the four order moments curve
from $\beta=0.7$ to $1.5$.}
\label{fig:compmfmg}
\end{figure}
\vspace{-2mm}
Using a diagrammatic method (details to be presented elsewhere) 
we were able to derive the first four
one-plaquette connected moments for both the vacuum and
anti-symmetric state. 
These moments immediately give analytic approximates for vacuum energy and
the mass gap.

Since the expected scaling behaviour for compact U$(1)_{2+1}$ is
of the form (\ref{scale}), the one-plaquette state results for the
plaquette expansion to first order are presented as a
$\mbox{ln}\left(M^{2}/\beta\right)$ versus $\beta$ plot in Figure
\ref{fig:compmfmg}. 
The results using one, three and four moments appear to be converging
towards the scaling slope. 
Clearly scaling is evident for
inverse-coupling values $\beta=0.7$ to $1.5$, passing the
transition point at $\beta=1$. The scaling form we find is given
by $M^{2}=\beta\,\mbox{exp}\left(-4.34\beta+5.34\right)$, and
agrees well with other estimates (as summarized in \cite{inprogress}).

\begin{figure}
\vspace{0mm}
\hbox{\hspace{0mm}\parindent=0mm
        \vtop{%
                \psfig{file=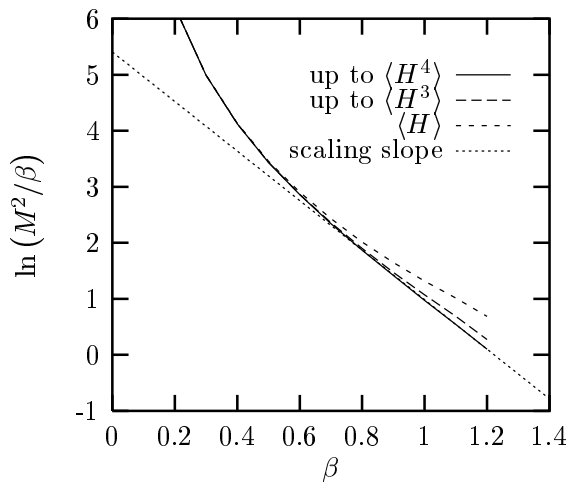,width=80mm}
                \vbox{\hsize=80mm} 
        }
}
\vspace{-2mm}
\caption{Convergence of the complete analytic mass gap results 
for one (variational), three and four one-plaquette
Hamiltonian moments 
to the continuum limit, using the series form for $\lambda(\beta)$.}
\label{fig:analytic}
\end{figure}

It should be noted that the minimization of $\langle H\rangle$ is the
only strictly numerical procedure (small as it is) in our calculation. 
A totally analytic estimation (keeping in line with the title of this paper)
is possible 
when one writes the function $\lambda(\beta)$ which minimizes $\langle H\rangle$
as a series expansion:
$\;\lambda=\beta^{2}/2-\beta^{6}/16+\ldots\,$.
Keeping the first five terms of the expansion (for which the resulting
values have converged), scaling is observed 
from $\beta=0.8$ to $1.2$~(Figure~\ref{fig:analytic}). 
The series breaks down at larger $\beta$ values. The scaling form is given
by $M^{2}=\beta\,\mbox{exp}\left(-4.42\beta+5.41\right)$, and again
agrees well with the numerical estimates available.     

In summary, applying the plaquette expansion we were able to identify scaling
of the anti-symmetric (or photon) mass gap in the weak-coupling
regime of compact U$(1)_{2+1}$ lattice gauge theory via a
completely analytic calculation utilizing only four one-plaquette
state Hamiltonian moments. Furthermore, although we have presented
results for only the first four moments (for which the calculation
is entirely analytic) it should in principle be possible to derive
higher moments with respect to this state. Indeed we have
preliminary results for the vacuum energy density to sixth
order\cite{inprogress} giving high accuracy over a large range of couplings.

\vspace{-2mm}
These results may have important implications for lattice gauge
theory -- the plaquette expansion method itself can be applied to
any extensive many-body problem. The only existing application of
the method to QCD$_{3+1}$ (including massless quarks) with respect
to the strong coupling state proved somewhat
disappointing\cite{holl94b} in that scaling of the
nucleon/meson mass ratio was not evident. However, the results presented here
indicate that the one-plaquette state may be of sufficient
quality to capture the essence of the continuum physics in lattice gauge theory at
relatively low orders. An exciting possibility is that the continuum limit
might finally be observed in the Hamiltonian formalism
for QCD using this method. 

To finish, we propose a semi-analytic program to study the 3+1 lattice
gauge theories in the Hamiltonian formalism. The main obstacle to
overcome is the calculation of the moments in 3+1 dimensions. The
appeal of 2+1 systems as test models is that the transformation
from link variables to plaquette variables is trivial and makes
the integrations tractable -- hence the analytic work reported in
this paper. In 3+1 dimensions it is well known that the
transformation cannot be carried out in closed form due to the
appearance of Biachi identities. However, the calculation of
moments in 3+1 dimensions by Monte-Carlo methods is actually a relatively
small scale numerical exercise. Because we are dealing with a
cluster expansion (we need only connected moments) the lattices
required fairly modest in extent, and are by definition only three
dimensional. The integrands become complicated as larger
correlations are included, which means that the statistics must be
very good, however, preliminary calculations have demonstrated
that the required precision is possible to achieve with quite modest
computing resources~\cite{inprogress}.

This work was supported by the Australian
Research Council and the Australian Postgraduate Award scheme.


\widetext

\begin{thebibliography}{10}
\bibitem{gopmack82}
M.G$\ddot{\mbox{o}}$pfert and G.Mack, {\it Commum. Math. Phys.} {\bf 82},  545
  (1982).

\bibitem{inprogress}
J.A.L.McIntosh, Analytic Lanczos Methodology and Scaling in
Hamiltonian Lattice Gauge Theory, Ph.D. thesis (2000).

\bibitem{kroeger}
E.B.Gregory, S.H.Guo, H.Kr\"{o}ger and X.Q.Luo, {\it Phys. Rev.} D {\bf 62}, 054508  (2000). 

\bibitem{kogsuss75}
J.Kogut and L.Susskind, {\it Phys. Rev.} D {\bf 11},  395  (1975).

\bibitem{holl93}
L.C.L.Hollenberg, {\it Phys. Rev.} D {\bf 47},  1640  (1993).

\bibitem{hornwein84}
D.Horn and M.Weinstein, {\it Phys. Rev.} D {\bf 30},  1256  (1984).

\bibitem{hwaf95}
L.C.L.Hollenberg and N.S.Witte, {\it Phys. Rev.} B {\bf 54},  16309  (1996).

\bibitem{hollwilwit95}
L.C.L.Hollenberg, M.Wilson and N.S.Witte, {\it Phys. Lett.} B {\bf 361},  81
  (1995).

\bibitem{macholl97}
J.A.L.McIntosh and L.C.L.Hollenberg, {\it Z. Phys.} C {\bf 76},  175  (1997).

\bibitem{lang84}
W.Langguth, {\it Z.Phys.} C {\bf 23},  289  (1984).

\bibitem{heysstump85}
D.W.Heys and D.R.Stump, {\it Nucl. Phys.} B {\bf 257},  19  (1985).

\bibitem{bish96}
S.J.Baker, R.F.Bishop and N.J.Davidson, {\it Phys. Rev.} D {\bf 53},  2610
  (1996).

\bibitem{bish97}
S.J.Baker, R.F.Bishop and N.J.Davidson, {\it Nucl. Phys.} B ({\it Proc.  Suppl.}) 
{\bf 53},  834  (1997).

\bibitem{dunp1}
A.Duncan and R.Roskies, {\it Phys. Rev.} D {\bf 31},  364  (1985).

\bibitem{duncan}
A.Duncan, {\it Nucl. Phys.} B {\bf 258},  125  (1985).

\bibitem{bish93}
R.F.Bishop, A.S.Kendall, L.Y.Wong and Y.Xian, {\it Phys. Rev.} D {\bf 48},  887
   (1993).

\bibitem{irvham84a}
A.C.Irving and C.J.Hamer, {\it Nucl. Phys.} B {\bf 230},  361  (1984); {\it Nucl. Phys.} B 
{\bf 235},  358  (1984).

\bibitem{dabring91}
A.Dabringhaus, M.L.Ristig and J.W.Clark, {\it Phys. Rev.} D {\bf 43},  1978
  (1991).

\bibitem{hamzheng93}
C.J.Hamer and Z.Weihong, {\it Phys. Rev.} D {\bf 48},  4435  (1993).

\bibitem{holl94b}
L.C.L.Hollenberg, {\it Phys. Rev.} D {\bf 50},  6917  (1994).

\end{thebibliography}
\end{document}